\documentclass[twocolumn,showpacs,aps,prb,superscriptaddress,floatfix,amsmath,amssymb]{revtex4}

\pdfoutput=1
\usepackage{graphicx}
\usepackage{dcolumn}
\usepackage{bm}
\usepackage{subfigure}

\begin{document}

\title{Muon Spin Rotation Measurements of Heterogeneous Field Response in Overdoped $\mathbf{ La_{2-x}Sr_{x}CuO_{4}}$}

\author{G. J. MacDougall}
\affiliation{Department of Physics and Astronomy, McMaster University,
Hamilton, Ontario, Canada, L8S-4M1}
\altaffiliation{Current address: Neutron Scattering Science Division, Oak Ridge National Laboratory, Oak Ridge, Tennessee 37831, USA
email: macdougallgj@ornl.gov}

\author{A. T. Savici}
\affiliation{Department of Physics, Columbia University,
New York, New York 10027, USA}
\affiliation{Department of Physics and Astronomy, Johns Hopkins University, Baltimore, Maryland 21218, USA}

\author{A. A. Aczel}
\affiliation{Department of Physics and Astronomy, McMaster University,
Hamilton, Ontario, Canada, L8S-4M1}

\author{R. J. Birgeneau}
\affiliation{Department of Physics, University of Toronto,
Toronto, Ontario, Canada, M5S-1A7}
\affiliation{Department of Physics and Lawrence Berkeley Laboratory, University of California, Berkeley, California 94720, USA}

\author{H. Kim}
\affiliation{Department of Physics, University of Toronto,
Toronto, Ontario, Canada, M5S-1A7}

\author{S.-J. Kim}
\affiliation{Department of Physics and Astronomy, McMaster University,
Hamilton, Ontario, Canada, L8S-4M1}

\author{T. Ito}
\affiliation{Department of Physics, Columbia University,
New York, New York 10027, USA}
\affiliation{National Institute of Advanced Industrial Science and Technology (AIST),
Tsukuba, Ibaraki 305-9562, Japan}

\author{J. A. Rodriguez}
\affiliation{Department of Physics and Astronomy, McMaster University,
Hamilton, Ontario, Canada, L8S-4M1}

\author{P. L. Russo}
\affiliation{Department of Physics, Columbia University,
New York, New York 10027, USA}

\author{Y. J. Uemura}
\affiliation{Department of Physics, Columbia University,
New York, New York 10027, USA}

\author{S. Wakimoto}
\affiliation{Department of Physics, University of Toronto,
Toronto, Ontario, Canada, M5S-1A7}
\affiliation{Quantum Beam Science Directorate, Japan Atomic Energy Agency, Tokai, Ibaraki, Japan 319-1195}

\author{C. R. Wiebe}
\affiliation{Department of Physics and Astronomy, McMaster University,
Hamilton, Ontario, Canada, L8S-4M1}
\affiliation{Department of Physics, Columbia University,
New York, New York 10027, USA}

\author{G. M.  Luke}
\affiliation{Department of Physics and Astronomy, McMaster University,
Hamilton, Ontario, Canada, L8S-4M1}
\affiliation{Canadian Institute for Advanced Research, Toronto, Ontario, Canada, M5G 1Z8}

\date{\today}

\begin{abstract}
Transverse-field muon spin rotation measurements of overdoped $\mathrm{La_{2-x}Sr_{x}CuO_{4}}$ reveal a large broadening of the local magnetic field distribution in response to applied field, persisting to high temperatures. The field-response is approximately Curie-Weiss like in temperature and is largest for the highest doping investigated. Such behaviour is contrary to the canonical Fermi-liquid picture commonly associated with the overdoped cuprates and implies extensive heterogeneity in this region of the phase diagram. A possible explanation for the result lies in regions of staggered magnetization about dopant cations, analogous to what is argued to exist in underdoped systems.
\end{abstract}


\maketitle

\section{Introduction}
Of the many recurring themes in cuprate research, one of the most persistent is atomic-scale heterogeneity. In hole-doped cuprates in particular, local probes have provided numerous examples where superconducting, electronic and magnetic properties vary on a lengthscale of tens of angstroms. Scanning tunneling spectroscopy (STS) regularly reports variations in the superconducting gap of $\mathrm{Bi_{2}Sr_{2}CaCu_{2}O_{8+\delta}}$ (BSCCO) which seem to correlate well with the location of dopant oxygen ions\cite{lang02,hoffman02,mcelroy05,mcelroy05_2,lee05}. Similar measurements on $\mathrm{La_{2-x}Sr_{x}CuO_{4}}$ (LSCO) reveal a spatial variation in the local density of states\cite{kato05}. A number of NMR and nuclear quadrupole resonance (NQR) studies have reported inequivalent sites and doping heterogeneity in La-based cuprate systems\cite{yoshimura89,tou92,haase00,haase02,singer02,singer05,williams05,graffe06,ofer06}.

Magnetic properties of underdoped systems are dominated by incommensurate order and fluctuations, commonly associated with one-dimensional variations in charge and spin density\cite{birgeneau06}. Zero-field muon spin rotation (ZF-$\mathrm{\mu SR}$) measurements of $\mathrm{La_{1.88}Sr_{0.12}CuO_{4}}$ and $\mathrm{La_{2}CuO_{4.11}}$ suggest that such heterogeneous `stripe order' can itself be fragmented, and that long-range order in these systems results from percolation between overlapping 15-30\AA~islands\cite{savici02}. Separate studies of $\mathrm{La_{1.85-y}Eu_{y}Sr_{0.15}CuO_{4}}$\cite{kojima03} and $\mathrm{La_{2-x}Sr_{x}CuO_{4+y}}$\cite{mohottala06} report similar phase separation over a wide range of doping, with superconducting and magnetic volumes mutually exclusive and oppositely correlated.

Upon doping $\mathrm{Zn^{2+}}$ ions onto the planar $\mathrm{Cu^{2+}}$ sites of $\mathrm{YBa_{2}Cu_{3}O_{6+\delta}}$ (YBCO), a number of NMR studies report the existence of staggered spins immediately about the substitution site\cite{alloul91,walstedt93,mahajan94,williams95,williams98,bobroff99,julien00,williams01,itoh03,ouazi04,ouazi06}. A local suppression of superconductivity in the same regions of the sample has been inferred from first $\mathrm{\mu SR}$\cite{nachumi96} and later STS\cite{pan00} studies, though specific interpretation of both NMR and $\mathrm{\mu SR}$ results is still a matter of debate\cite{williams_comment,bernhard98_2}. Recent theoretical work suggests that similar antiferromagnetic islands are expected in underdoped LSCO around dopant $\mathrm{Sr^{2+}}$ ions due to correlation effects, and that this may explain several properties of the La-based cuprate superconductors\cite{andersen07}.

Whereas heterogeneity is well-established in the underdoped cuprates, the situation is much less clear on the overdoped side of the phase diagram. It is widely believed that the overdoped materials behave like canonical Fermi-liquids, and this belief is supported by select transport measurements\cite{proust02,nakamae03,kaminksi03,plate05,hussey03}. Effects from $\mathrm{Zn^{2+}}$ impurities in YBCO are also less pronounced beyond optimal doping\cite{alloul91,mahajan94,bobroff99,williams95}, and even glassy order disappears around $x$=0.19 in LSCO\cite{panagopolous02}. On the other hand, bulk susceptibility studies of both LSCO and $\mathrm{Tl_{2}Ba_{2}CuO_{6+\delta}}$ (Tl-1201) observe a Curie component emerging with progressive overdoping\cite{kubo91,oda91,nakano94,wakimoto05}, in stark contrast to the temperature independent Pauli susceptibility expected for a Fermi liquid. Moreover, STS on heavily overdoped $\mathrm{Bi_{2-x}Pb_{x}Sr_{2}CuO_{y}}$ reports nanoscale variations in the superconducting gap\cite{mashima06}, unpaired carriers have been observed at low temperatures in overdoped LSCO\cite{wang07}, and real space phase separation has been argued indirectly from $\mathrm{\mu SR}$\cite{uemura93}, optical conductivity\cite{schutzmann94,uchida96}, magnetization\cite{wen00} and neutron scattering\cite{wakimoto04,wakimoto05} measurements.

In recent years, a handful of transverse-field (TF) $\mathrm{\mu SR}$ studies\cite{savici05,ishida07,sonier08} have reported an heterogeneous magnetic field response in single crystals of underdoped LSCO\cite{savici05,ishida07,sonier08} and YBCO\cite{sonier08}, in the form of a symmetric broadening of the local field distribution in response to applied field. First reported by Savici \emph{et al.} in crystals containing fragmented stripe order\cite{savici05}, the effect was originally associated with the staggered susceptibility of the heterogeneous spin system. Data from Ishida \emph{et al.}\cite{ishida07} at a lower doping were interpreted in a similar way. Most recently, however, Sonier \emph{et al.}\cite{sonier08} demonstrated that the size of the field-induced broadening closely correlates with superconducting $\mathrm{T_{c}}$ in underdoped YBCO, and inferred local patches of superconductivity at temperatures well-above the bulk transition. 

In this paper, we address the magnetic field response of overdoped LSCO with TF-$\mathrm{\mu SR}$. Preliminary work by us suggests a significant response to applied field at very high dopings\cite{macdougall06}. Here, we present data on four crystals ranging from optimal to heavily overdoped in a variety of applied fields and temperatures, comparing directly to previous work. We show that in every crystal investigated there exists a spatially heterogeneous response to applied field which increases with decreasing temperature in a Curie-Weiss-like fashion. The dominant trend is a linear increase of field response with $x$ through the origin, with a notable change in temperature dependence above optimal doping. In $\mathrm{La_{1.7}Sr_{0.3}CuO_{4}}$, we further show that this response is highly anisotropic, being larger when the field is applied perpendicular to the copper-oxide planes. Several possible pictures are put forth, and the ability of each to explain our data is discussed.

\section{Experimental Details and Results}
Four crystals of LSCO were grown via the traveling solvent floating zone method. Crystals with $x$=0.15 and $x$=0.19 were grown at AIST Tsukuba, and crystals with $x$=0.25 and $x$=0.30 were grown at the University of Toronto. Samples with $x$=0.19, $x$=0.25 and $x$=0.3 were cut into thin plates with the $\hat{c}$-axis perpendicular to the large face. An additional crystal with $x$=0.30 was cut with the $\hat{c}$-axis in the plane of the plate to test for signal anisotropy. The crystal with $x$=0.15 was a half cylinder with approximate radius of $\sim$5mm. Crystals for the larger two dopings are from the same batch as those studied earlier by Wakimoto \emph{et al.} with neutron scattering\cite{wakimoto04}. SQUID magnetization measurements on these two samples confirmed results by Wakimoto on related powder samples\cite{wakimoto05}, where a Curie component grows with overdoping by an amount equivalent to 0.5$\mu _{B}$ per additional hole.

TF-$\mathrm{\mu SR}$\cite{summerschool} data were collected using the HI-Time spectrometer on the M-15 beamline of TRIUMF in Vancouver, Canada. Fields ranging from 0.2T to 7T were applied perpendicular to the large face of the crystals, equivalent to the crystalline $\hat{c}$-axis for all but one sample. Applied fields were measured independently by simultaneously observing the precession of muons landing in an adjacent crystal of $\mathrm{Ca(CO_{3})}$.

\begin{figure}[tpb]
\centering
\subfigure{
   \includegraphics[width=0.45\columnwidth]{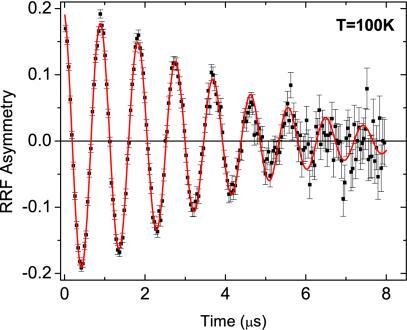}}
\subfigure{
   \includegraphics[width=0.45\columnwidth]{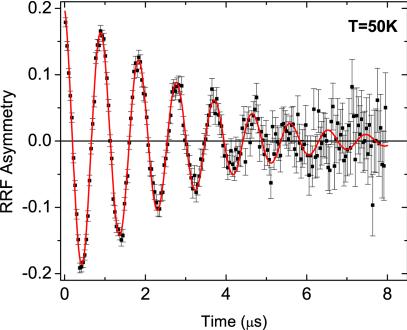}}
\subfigure{
   \includegraphics[width=0.45\columnwidth]{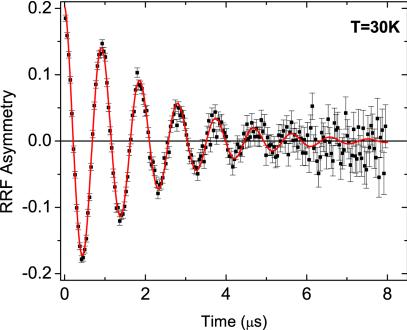}}
\subfigure{
   \includegraphics[width=0.45\columnwidth]{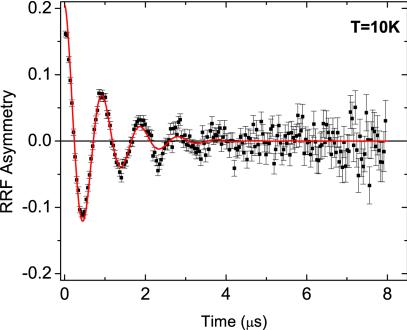}}
\subfigure{
   \includegraphics[width=0.45\columnwidth]{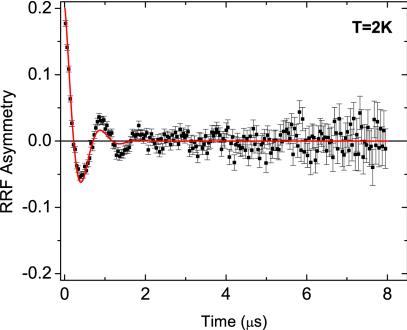}}
\subfigure{
   \includegraphics[width=0.45\columnwidth]{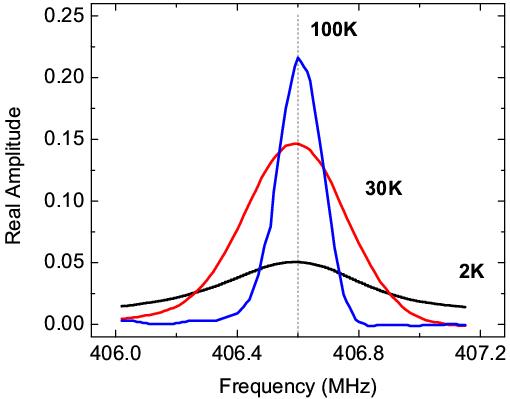}}
\caption{\label{fig:spectra}(a)-(e)RRF asymmetries resulting from $TF-\mu SR$ measurements of single crystal $\mathrm{La_{1.7}Sr_{0.3}CuO_{4}}$ in an applied field of $\mu_{0}$H=3T at several temperatures. Field was applied perpendicular to the copper-oxide planes. (f)FFTs of three time spectra. The vertical dashed line shows the location of the applied field.}
\end{figure}

Figures ~\ref{fig:spectra}(a)-(e) show sample $\mathrm{\mu SR}$ spectra from $\mathrm{La_{1.7}Sr_{0.3}CuO_{4}}$ in an applied field of $\mu_{0}$\textbf{H}=3T $\| \hat{c}$ at five temperatures between 100 and 2K. A dramatic relaxation of our TF-$\mathrm{\mu SR}$ asymmetry is evident, with the rate of relaxation increasing more rapidly as the temperature is lowered. This is in contrast to separate measurements of the same crystal in zero field\cite{macdougall08}, which shows no appreciable temperature dependence down to 10K. In the absence of magnetic fluctuations, the TF-$\mathrm{\mu SR}$ asymmetry function is proportional to the cosine Fourier transform of the local distribution of fields at the muon site\cite{summerschool}, and thus the relaxation rates of the time spectra are a measure of the width of the local field distribution at the muon site. This is demonstrated explicitly in Fig.~\ref{fig:spectra}(f), which shows fast Fourier transforms of three time spectra. Of particular note is lack of net $\textit{shift}$ of the average field from the applied field (dotted line), despite a significant increase in the $\textit{width}$ of the distribution. Similar spectra were seen for every crystal investigated and for every (non-zero) applied field.

In order to quantify the magnetic response, we fit all spectra to the form:
\begin{equation}
A(t)=A_{0}\cos(\omega _{\mu} t+\phi)e^{-\sigma^{2} t^{2}}e^{-\lambda t},
\label{eq:asy}
\end{equation}
where $\omega _{\mu}$ is the mean precession frequency of the muons ($\propto$ average local field), and $\lambda$ is the rate of field-induced relaxation. The Gaussian term in Eq.~(\ref{eq:asy}) is present to account for finite field-width due to randomly oriented nuclear dipole moments and applied field heterogeneity. It is temperature independent and was held constant to a value chosen by separate calibration measurements. The exponential form for the field response is equivalent to a Lorentzian-like broadening of the field distribution and was chosen to allow direct comparison of the current work to previous studies\cite{savici05,ishida07,sonier08}. Careful examination of the $\mathrm{\mu SR}$ lineshape reveals that spectra are better described by a root exponential function at low temperatures, but all spectra were well-described by Eq.~(\ref{eq:asy}) except when $x$=0.30 and T$<$10K.

\begin{figure}[thb]
\centering
\subfigure{ \includegraphics[width=0.45\columnwidth]{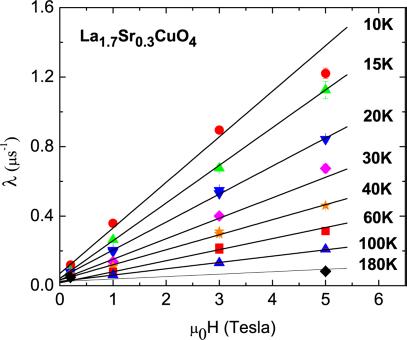}}
\subfigure{ \includegraphics[width=0.45\columnwidth]{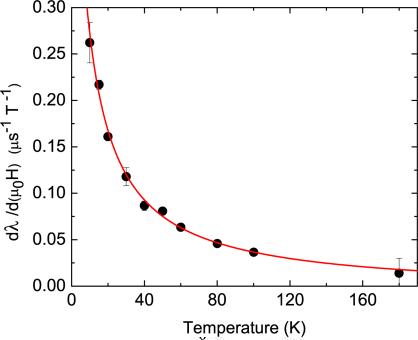}}
\subfigure{ \includegraphics[width=0.45\columnwidth]{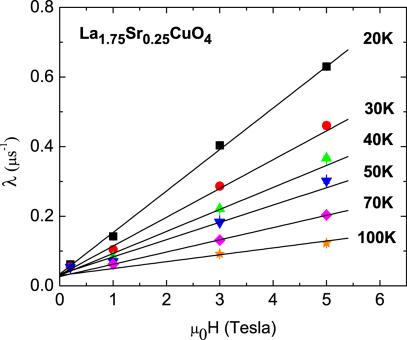}}
\subfigure{ \includegraphics[width=0.45\columnwidth]{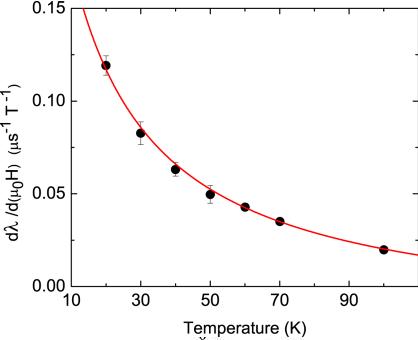}}
\subfigure{ \includegraphics[width=0.45\columnwidth]{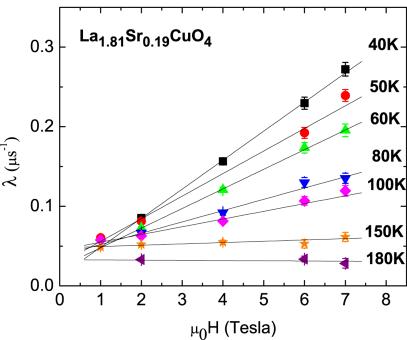}}
\subfigure{ \includegraphics[width=0.45\columnwidth]{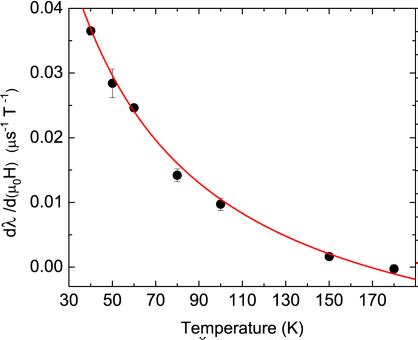}}
\subfigure{ \includegraphics[width=0.45\columnwidth]{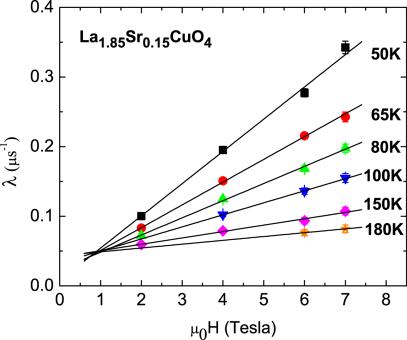}}
\subfigure{ \includegraphics[width=0.45\columnwidth]{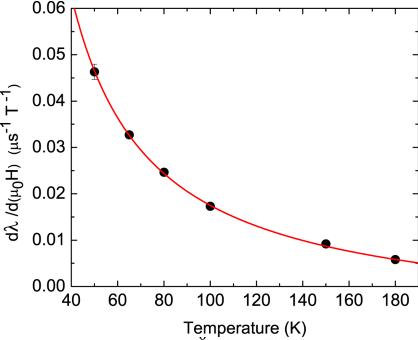}}
\caption{\label{fig:relaxations}(Left) The field dependence of the $\mathrm{\mu SR}$ linewidth for each of the four crystals, extracted from Eq.~\ref{eq:asy}. (Right) The slopes of the $\frac{d\lambda}{dH}$ curves as a function of temperature. Solid lines in both panels represent fits described in the text. }
\end{figure}

Figure~\ref{fig:relaxations} shows the extracted values of $\lambda$ for each of the four crystals as a function of applied field and temperature. Relaxation rates are largest for our highest doping, and generally get smaller with decreasing doping- though the $x$=0.15 sample shows a slightly larger effect than the $x$=0.19 sample. At all temperatures, the rates are roughly linear in field and extrapolate to small positive values, revealing a slight overestimation of $\sigma$ in Eq.~(\ref{eq:asy}). Slopes of the $\lambda$ vs. $\mu_{0}$H curves increase gradually as the samples were cooled, with an approximate Curie-Weiss-like temperature dependence.

The temperature dependence for each crystal was fit to the form:
\begin{equation}
\frac{1}{\mu_{0}}\frac{d\lambda}{dH}(T,x) = \frac{C(x)}{T-\Theta(x)}+B(x),
\label{eq:curie}
\end{equation}
where B($x$) was included to account for possible temperature independent sources of relaxation, such as Pauli susceptibility from a heterogeneous metallic state, Van-Vleck susceptibility from randomly placed dopant cations, and poor estimation of the applied field heterogeneity in Eq.~\ref{eq:asy}. Curves of best fit are shown as solid lines in Fig.~\ref{fig:relaxations}. Though Eq.~\ref{eq:curie} provides a useful characterization of the current data, it is important to note that other forms for the temperature dependence describe the data equally well.

\begin{figure}[t]
\centering
\includegraphics[width=0.9\columnwidth]{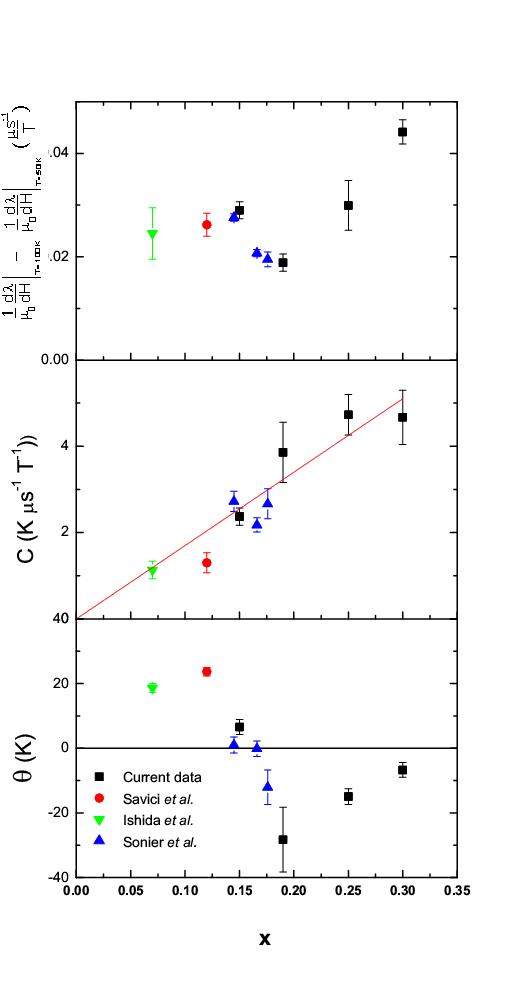}
\caption{\label{fig:xdependence}The doping evolution of the heterogeneous field-response. The top panel shows the increase of the absolute value of $\frac{1}{\mu_{0}}\frac{d\lambda}{dH}$ at T=50K over that at T=100K. The following two panels show values of $C$ and $\Theta$, extracted from fits to Eq.~\ref{eq:curie}. The line in the centre panel is a guide to the eye.}
\end{figure}

The doping dependence of the heterogeneous field response is summarized in Fig.~\ref{fig:xdependence}. In addition to the data presented above, we have also included data previously published in Ref.~\onlinecite{savici05} ($x$=0.12), and data extracted from Refs.~\onlinecite{ishida07} ($x$=0.07) and \onlinecite{sonier08} ($x$=0.145, 0.166 and 0.176). The topmost panel shows the increase in $\frac{1}{\mu_{0}}\frac{d\lambda}{dH}$ at T=50K over that at T=100K. The current data set is consistent with the work of Sonier \emph{et al.} over the same doping range. The overall evolution of the field response in the underdoped materials is also qualitatively consistent with the trend reported by Sonier in YBCO, where $\lambda$ varies in a way that reflects superconducting $T_{c}$. Above optimal doping however, the system breaks from this trend, and the relaxation increases rapidly with $\mathrm{Sr^{2+}}$ content up to the largest doping explored. The middle and lower panels of Fig.~\ref{fig:xdependence} show the respective values of $C(x)$ and $\Theta (x)$ extracted from fits to Eq.~\ref{eq:curie}. Fits of data from Refs.~\onlinecite{ishida07} and \onlinecite{sonier08} were performed assuming a linear field dependence through the origin, which neglects field-independent sources of relaxation. Though there is some scatter, the dominant trend for $C(x)$ is a near linear increase with $x$, passing through the origin at zero doping. Fit values for $\Theta (x)$, large and positive in the underdoped materials, go negative at slight overdoping and tend towards zero as $x$ get large. We note that this critical doping has in the past been associated with the disappearance of glassy order in ZF-$\mathrm{\mu SR}$ measurements\cite{panagopolous02} and the emergence of Curie term in the bulk magnetization\cite{kubo91,oda91,nakano94,wakimoto05}. Specific interpretation of these trends will be discussed below.

\begin{figure}[thb]
\centering
\subfigure{\includegraphics[width=0.45\columnwidth]{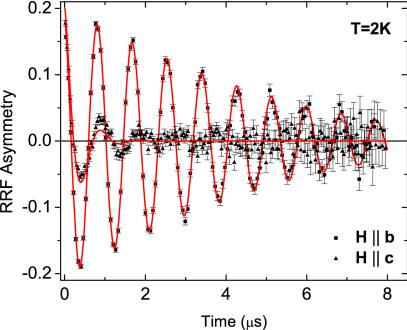}}
\subfigure{\includegraphics[width=0.45\columnwidth]{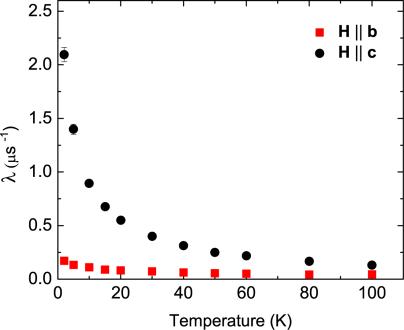}}
\caption{\label{fig:anisotropy}(a)Raw asymmetry from $TF-\mu SR$ measurements of $\mathrm{La_{1.7}Sr_{0.3}CuO_{4}}$ at T=2K and $\mu_{0}$H=3T, applied perpendicular and parallel to the copper-oxide planes. (b)The relaxation rates resulting from fitting analagous spectra as that in (a), but over a range of temperatures.} \end{figure}

Additional measurements were made with a separate crystal of $\mathrm{La_{1.7}Sr_{0.3}CuO_{4}}$ with $\textbf{H} \| \hat{b}$. The main result is demonstrated in Fig.~\ref{fig:anisotropy}. The left panel compares the spectra at T=2K, with a field of $\mu_{0}$H=3T applied parallel and perpendicular to the copper-oxide planes. The right panel shows the result of fitting such spectra with Eq.~\ref{eq:asy} at several temperatures. A large signal anisotropy is immediately apparent, with the heterogeneous field response an order of magnitude larger when the field is applied perpendicular to the planes. This anisotropy is significantly larger than that observed by Savici in $\mathrm{La_{1.875}Ba_{0.125}CuO_{4}}$\cite{savici05}.

\section{Discussion}

The present data was taken in the `transverse-field geometry', with the field applied perpendicular to the initial muon spin direction\cite{summerschool}. Previously published data on both underdoped\cite{savici05} and overdoped\cite{macdougall06} systems reveal zero relaxation within detection limits when in the `longitudinal-field geometry', where the field is parallel to the initial muon spin direction. As discussed in detail by Savici\cite{savici05}, these longitudinal-field $\mathrm{\mu SR}$ experiments put upper limits on possible transverse fluctuation rates associated with the observed magnetism. This observation allows us to confidently associate relaxation of the TF-$\mathrm{\mu SR}$ asymmetry with a broadening of the static internal magnetic field distribution, which has been an implicit assumption in the discussion thus far.

It is important to note that this broadening of the internal field distribution is largely without an accompanying shift. By comparing the precession frequency of muons in LSCO to those in a reference crystal of $\mathrm{Ca(CO_{3})}$, we observe that the average field of the muon site is within 0.5mT of the applied field at all temperatures in this study and over the entire doping range. This corresponds to a shift of $<$0.01$\%$ for $\mu_{0}$\textbf{H}=5T. In contrast, the distribution of fields in $\mathrm{La_{1.7}Sr_{0.3}CuO_{4}}$ has a FWHM of 5mT at T=2K in the same applied field. Stated simply, muons at different (crystallographically equivalent) sites are observing radically different local fields at low temperatures, even though the average properties are changing very little. A large field-width with small average shift is contrary to what one would expect for a uniform sample response or commensurate long-range order. In fact, quite independent of model, this observation implies extensive heterogeneity in the local magnetic ground state. This comes as quite a surprise in systems which are often touted as canonical Fermi liquids. It is only appropriate then to speculate on possible interpretations for the present data, and the remainder of the paper will discuss the feasibility of several such models.

NQR studies have successfully applied a distribution-of-hole model to explain line-broadening in underdoped cuprates\cite{singer02,haase00}. In such a picture, holes are quasi-localized in the copper-oxide planes, and each frequency in a given distribution is associated with a unique hole density. Though overdoped systems are expected to be more metallic (and thus have more delocalized charge carriers), it is not unreasonable to expect that the holes in the current samples are likewise distributed. Indeed, this idea is seemingly supported by earlier SQUID characterization, which revealed a small ($<5\%$) superconducting volume with a $T_{c} \sim$ 30K in some overdoped samples\cite{wakimoto04}. However, the shifts in $\mathrm{\mu SR}$ measurements of LSCO are orders of magnitude smaller than those observed by NQR, and the sheer magnitude of the present effect makes it difficult to explain by a variation of hole densities alone. For example, muons in $\mathrm{La_{1.7}Sr_{0.3}CuO_{4}}$ observe fields \emph{on average} 5mT removed from the applied field at low temperature, with many muons seeing even larger fields. Yet, there is no doping of LSCO which would correspond to a 5mT shift. Said another way, there is simply not enough variation in shift across the entire LSCO phase diagram to explain the distribution of the fields in $\mathrm{La_{1.7}Sr_{0.3}CuO_{4}}$. More fundamentally, this model is incomplete, as it does not explain the doping evolution of magnetic properties nor the source of the doping heterogeneity; it simply notes the \emph{existence} of heterogeneity. Thus, for a deeper understanding of these systems, further discussion of magnetic response is required.

As mentioned in Sec. I, a similar response to applied field was observed by Sonier \emph{et al.} in crystals of LSCO and YBCO, and successfully modeled as a consequence of heterogeneous superconductivity persisting to high temperatures\cite{sonier08}. This model was predicated on the observation that the rate of relaxation in the underdoped YBCO systems was proportional to superconducting $\mathrm{T_{c}}$, and indeed the collective data in underdoped LSCO appear to follow a similar trend. However, the field response for the three most heavily doped samples increases monotonically with Sr concentration, despite a weakening of superconductivity over the same doping range. Thus, superconductivity seems like a less suitable candidate to explain a heterogeneous field-response in overdoped systems, and implies at least the existence of a separate relaxation mechanism which dominates at for large values of $x$.

Other $\mathrm{\mu SR}$ data have been interpreted as a signature of spin magnetism\cite{savici05,ishida07}. Again, heavily overdoped cuprates are known to be free from magnetic order or glassiness and are widely expected to be non-magnetic. Yet, the Curie term in bulk magnetization data is suggestive of emergent local moments on the overdoped side of the phase diagram\cite{kubo91,oda91,nakano94,wakimoto05}, and there are theoretical grounds to expect such behaviour\cite{eskes88, eskes91}. It is commonly accepted that the most favourable band for doped holes in underdoped systems is the so-called Zhang-Rice band\cite{zhang88}- a singlet band formed from the hybridization of $\mathrm{Cu~3d_{x^{2}-y^{2}}}$ with $\mathrm{O~2p_{x}}$ and $\mathrm{O~2p_{y}}$ orbitals. However, long-standing studies by Eskes and Sawatzky\cite{eskes88, eskes91} predict a triplet band of mostly $\mathrm{Cu~3d_{3z^{2}-r^{2}}}$ character at only slightly higher energies. This triplet band is predicted to become more stable with decreasing apical oxygen distance, which can be affected by the presence of dopant cations\cite{eskes88, eskes91}. There are existing x-ray studies that support both a decreasing apical oxygen distance\cite{maignan90, martin95, haskel97} and increased occupation of the $\mathrm{Cu~3d_{3z^{2}-r^{2}}}$ orbital\cite{chen92, pellegrin93, srivastava96} in overdoped systems.

The slowing of longitudinal fluctuations amongst such emergent moments with applied field could certainly broaden the local magnetic field distribution in a similar manner to what is seen. To test the ability of this picture to explain our data, we simulated the expected $\mathrm{\mu SR}$ asymmetry function from dipole coupling to a system 0.5$\mu_{B}$ aligned moments on 8$\%$ of the planar copper sites. The density and size of the magnetic moments were entirely constrained by bulk magnetization data of Wakimoto \emph{et al.}\cite{wakimoto05} for $\mathrm{La_{1.7}Sr_{0.3}CuO_{4}}$. As expected for a dilute moment system\cite{summerschool}, the simulated asymmetry decayed exponentially with time. However, even for completely saturated moments, the maximum expected relaxation rate in such a picture is $\sim$1.5$\mu s^{-1}$- about 50$\%$ smaller than the largest rate seen in $\mathrm{La_{1.7}Sr_{0.3}CuO_{4}}$. More fundamentally, simulations involving isolated and unconstrained (also from bulk magnetization) moments always result in a completely isotropic distribution of fields, in stark contrast to the large signal anisotropy demonstrated in Fig.~\ref{fig:anisotropy}. One may be able to reconcile an isolated moment picture with the data, but only if one includes a(n) (anisotropic) staggered sample response which at least partially compensates a larger local moment.

Alternatively, Kopp \emph{et al.}\cite{kopp07} have argued that the overdoped cuprates may be characterized by competing ferromagnetism, with an eventual transition to a ferromagnetic ground state at some critical doping. In support of this idea, band calculations by Barbiellini and Jarlborg\cite{barbiellini08} predict that regions of weak ferromagnetism should emerge around clusters of dopant ions in $\mathrm{La_{2-x}Ba_{x}CuO_{4}}$ at high dilution. As with the isolated moment picture however, it is difficult to reproduce the largest observed relaxation rates within a ferromagnetic cluster picture, since now the \emph{total} moment in the system is constrained by bulk magnetization to be 0.5$\mathrm{\mu_{B}}$ per overdoped hole. Specific modeling suggests the ferromagnetic cluster model can increase the predicted relaxation rate by up to 20$\%$ over the isolated moment picture, but one is still unable to recreate the size or anisotropy of the field-response observed in $\mathrm{La_{1.7}Sr_{0.3}CuO_{4}}$ without including a staggered component to the local magnetism.

Perhaps a simpler explanation of the observed line-broadening is that it reflects clusters of staggered magnetic moments, possibly seeded by the dopant cations themselves. The temperature, field and orientational dependence of the current effect bear striking resemblance to data previously reported by NMR\cite{alloul91,walstedt93,mahajan94,williams95,williams98,bobroff99,julien00,williams01,itoh03,ouazi04,ouazi06} and $\mathrm{\mu SR}$\cite{mendels94,bernhard98} studies of Zn-doped YBCO, from which many infer the existence of staggered islands. Isolated islands with a net magnetic moment would act much like isolated paramagnetic moments, and would broaden the local field distribution in similar way. Yet, the net moment of an island would be greatly reduced from the total local moment and could easily reconcile the small effect seen in bulk magnetization with the larger effect seen here (c.f. Ref.~\onlinecite{mendels99}). Specific simulations by us confirm that the current data for $x$=0.30 are consistent with an island with radius $r \leq$ one lattice parameter. In such a scenario, the linear increase of $C(x)$ with $x$ in Fig.~\ref{fig:xdependence} is immediately understood to be a consequence of an increasing number of nucleation centers for impurity-induced magnetism. The rapid decrease in $\Theta (x)$ upon overdoping might reflect a shrinking magnetic correlation length (and thus island size), which would increase the total uncompensated spin of the islands and enhance $\bf{q}$=0 magnetic susceptibility at the expense of the staggered susceptibility. Consistent with shrinking magnetic correlations, we again mention the disappearance of glassy static magnetism\cite{panagopolous02} and the emergence of Curie-like magnetization\cite{kubo91,oda91,nakano94,wakimoto05} around the same doping. The enhanced field-response of $\mathrm{La_{1.75}Eu_{0.1}Sr_{0.15}CuO_{4}}$ over LSCO\cite{savici05} and $\mathrm{Y_{0.8}Ca_{0.2}Ba_{2}Cu_{3}O_{6}}$ over YBCO\cite{bernhard98} further supports the central role of dopant cations rather than doped holes.

The interpretation of NMR results in Zn-doped YBCO is still being actively discussed\cite{williams_comment}. However, local regions of staggered magnetization about dopant $\mathrm{Zn^{2+}}$ ions are predicted by a number of theoretical studies and expected to arise from correlation effects, even in metallic cuprates\cite{gabay07}. Andersen \emph{et al.} recently argued that completely analogous physics is expected in Zn-free LSCO due to weak Coulombic disorder from $\mathrm{Sr^{2+}}$ cations\cite{andersen07}. Thus, it is perhaps not surprising that we have observed data consistent with staggered moments. Whether it is appropriate to apply these underdoped studies to the more Fermi-liquid-like overdoped systems is not obvious and deserves further study. In support of this idea however, we note that strong correlations have already been inferred from an enhanced Kadowaki-Woods ratio in $\mathrm{La_{1.7}Sr_{0.3}CuO_{4}}$\cite{hussey03}, and both inelastic neutron scattering\cite{wakimoto05} and $\mathrm{\mu SR}$\cite{risdiana08} report an enhancement of antiferromagnetic fluctuations with Zn-doping in heavily overdoped systems. Antiferromagnetic fluctuations seem to decrease with Sr-doping in the Zn-free (and Zn-doped) systems\cite{wakimoto04}, but this may actually reflect a decreasing magnetic correlation length, inferred above to be less than one lattice parameter at $x$=0.30. Further, the application of magnetic field may encourage staggered magnetism where little is seen in zero field. Indeed, the central role of field is noted in theoretical studies of staggered magnetism in Zn-doped LSCO\cite{gabay07}, and in at least one cuprate system, applied field is seen to modify neutron spectra in a way that parallels disorder\cite{lee04}. To explore these ideas more thoroughly, we strongly encourage future measurements of the inelastic neutron spectrum of heavily overdoped cuprate systems in applied field and with varying levels of disorder.

\section{Summary and Conclusions}
We have presented TF-$\mathrm{\mu SR}$ data on overdoped $\mathrm{La_{2-x}Sr_{x}CuO_{4}}$ which shows a large broadening of the local magnetic field distribution in response to applied field over a range of temperatures. The temperature, field and orientational dependence closely reflects similar data reported in underdoped materials. Characterization of the data through fits to a Curie-Weiss function shows a roughly linear increase of $C(x)$ with $\mathrm{Sr^{2+}}$ concentration, but a marked shift in $\Theta (x)$ above optimal doping, where a Curie term in the bulk magnetization also emerges.

Though emergent paramagnetic moments and ferromagnetic clusters have been explored as possible explanations for the current data, it is difficult to reconcile the large local response with the weak bulk magnetization without including a staggered component to the local magnetism. We suggest that a simpler explanation of our data is dopant-induced staggered magnetization seeded by $\mathrm{Sr^{2+}}$ disorder and stabilized by field. Future experiments to investigate the effects of field and disorder in the overdoped cuprates are encouraged.

We emphasize the role that heterogeneity must play in any explanation of the current data. This is in contrast to the homogeneous Fermi-liquid ground state envisioned by many to exist in these systems, and implies at least strong correlation effects in the overdoped region of the phase diagram. The heterogeneous field response seen here complements previous reports of phase separation at low temperatures, and suggests that heterogeneous superconductivity at low temperatures may have its origin in a heterogeneous metallic state above $\mathrm{T_{c}}$.

\section{Acknowledgements}
The authors would like to acknowledge useful conversations with J.E. Sonier and G.A. Sawatzsky. Work at McMaster was supported by NSERC and CIFAR. Work at Columbia was supported by NSF under DMR-05-02706 and DMR-08-06846. The work at Lawrence Berkeley Laboratory was supported by the Office of Basic energy Sciences, U.S. Dept. of Energy under Contract No. DE-AC03-76SF0098. We appreciate the hospitality and technical assistance of the TRIUMF Centre for Molecular and Materials Science, where these experiments were performed.


\begin{thebibliography}{74}
\expandafter\ifx\csname natexlab\endcsname\relax\def\natexlab#1{#1}\fi
\expandafter\ifx\csname bibnamefont\endcsname\relax
  \def\bibnamefont#1{#1}\fi
\expandafter\ifx\csname bibfnamefont\endcsname\relax
  \def\bibfnamefont#1{#1}\fi
\expandafter\ifx\csname citenamefont\endcsname\relax
  \def\citenamefont#1{#1}\fi
\expandafter\ifx\csname url\endcsname\relax
  \def\url#1{\texttt{#1}}\fi
\expandafter\ifx\csname urlprefix\endcsname\relax\def\urlprefix{URL }\fi
\providecommand{\bibinfo}[2]{#2}
\providecommand{\eprint}[2][]{\url{#2}}

\bibitem[{\citenamefont{Lang et~al.}(2002)}]{lang02}
\bibinfo{author}{\bibfnamefont{K.~M.} \bibnamefont{Lang}} \bibnamefont{et~al.},
  \bibinfo{journal}{Nature} \textbf{\bibinfo{volume}{415}},
  \bibinfo{pages}{412} (\bibinfo{year}{2002}).

\bibitem[{\citenamefont{Hoffman et~al.}(2002)}]{hoffman02}
\bibinfo{author}{\bibfnamefont{J.~E.} \bibnamefont{Hoffman}}
  \bibnamefont{et~al.}, \bibinfo{journal}{Science}
  \textbf{\bibinfo{volume}{295}}, \bibinfo{pages}{466} (\bibinfo{year}{2002}).

\bibitem[{\citenamefont{McElroy et~al.}(2005{\natexlab{a}})}]{mcelroy05}
\bibinfo{author}{\bibfnamefont{K.}~\bibnamefont{McElroy}} \bibnamefont{et~al.},
  \bibinfo{journal}{Phys.\ Rev.\ Lett.} \textbf{\bibinfo{volume}{94}},
  \bibinfo{pages}{197005} (\bibinfo{year}{2005}{\natexlab{a}}).

\bibitem[{\citenamefont{McElroy et~al.}(2005{\natexlab{b}})}]{mcelroy05_2}
\bibinfo{author}{\bibfnamefont{K.}~\bibnamefont{McElroy}} \bibnamefont{et~al.},
  \bibinfo{journal}{Science} \textbf{\bibinfo{volume}{309}},
  \bibinfo{pages}{1049} (\bibinfo{year}{2005}{\natexlab{b}}).

\bibitem[{\citenamefont{Lee et~al.}(2005)}]{lee05}
\bibinfo{author}{\bibfnamefont{J.}~\bibnamefont{Lee}} \bibnamefont{et~al.},
  \bibinfo{journal}{J. Phys. Chem. Solids} \textbf{\bibinfo{volume}{66}},
  \bibinfo{pages}{1370} (\bibinfo{year}{2005}).

\bibitem[{\citenamefont{Kato et~al.}(2005)}]{kato05}
\bibinfo{author}{\bibfnamefont{T.}~\bibnamefont{Kato}} \bibnamefont{et~al.},
  \bibinfo{journal}{Phys.\ Rev.\ B} \textbf{\bibinfo{volume}{72}},
  \bibinfo{pages}{144518} (\bibinfo{year}{2005}).

\bibitem[{\citenamefont{Yoshimura et~al.}(1989)}]{yoshimura89}
\bibinfo{author}{\bibfnamefont{K.}~\bibnamefont{Yoshimura}}
  \bibnamefont{et~al.}, \bibinfo{journal}{J.\ Phys.\ Soc.\ Jap.}
  \textbf{\bibinfo{volume}{58}}, \bibinfo{pages}{3057} (\bibinfo{year}{1989}).

\bibitem[{\citenamefont{Tou et~al.}(1992)}]{tou92}
\bibinfo{author}{\bibfnamefont{H.}~\bibnamefont{Tou}} \bibnamefont{et~al.},
  \bibinfo{journal}{J.\ Phys.\ Soc.\ Jap.} \textbf{\bibinfo{volume}{61}},
  \bibinfo{pages}{1477} (\bibinfo{year}{1992}).

\bibitem[{\citenamefont{Haase et~al.}(2000)}]{haase00}
\bibinfo{author}{\bibfnamefont{J.}~\bibnamefont{Haase}} \bibnamefont{et~al.},
  \bibinfo{journal}{J.\ Supercond.} \textbf{\bibinfo{volume}{13}},
  \bibinfo{pages}{723} (\bibinfo{year}{2000}).

\bibitem[{\citenamefont{Haase et~al.}(2002)}]{haase02}
\bibinfo{author}{\bibfnamefont{J.}~\bibnamefont{Haase}} \bibnamefont{et~al.},
  \bibinfo{journal}{J.\ Supercond.} \textbf{\bibinfo{volume}{15}},
  \bibinfo{pages}{339} (\bibinfo{year}{2002}).

\bibitem[{\citenamefont{Singer et~al.}(2002)}]{singer02}
\bibinfo{author}{\bibfnamefont{P.~M.} \bibnamefont{Singer}}
  \bibnamefont{et~al.}, \bibinfo{journal}{Phys.\ Rev.\ Lett.}
  \textbf{\bibinfo{volume}{88}}, \bibinfo{pages}{047602}
  (\bibinfo{year}{2002}).

\bibitem[{\citenamefont{Singer et~al.}(2005)}]{singer05}
\bibinfo{author}{\bibfnamefont{P.}~\bibnamefont{Singer}} \bibnamefont{et~al.},
  \bibinfo{journal}{Phys.\ Rev.\ B} \textbf{\bibinfo{volume}{72}},
  \bibinfo{pages}{014537} (\bibinfo{year}{2005}).

\bibitem[{\citenamefont{Williams et~al.}(2005{\natexlab{a}})}]{williams05}
\bibinfo{author}{\bibfnamefont{G.}~\bibnamefont{Williams}}
  \bibnamefont{et~al.}, \bibinfo{journal}{Phys.\ Rev.\ B}
  \textbf{\bibinfo{volume}{71}}, \bibinfo{pages}{014515}
  (\bibinfo{year}{2005}{\natexlab{a}}).

\bibitem[{\citenamefont{Graffe et~al.}(2006)}]{graffe06}
\bibinfo{author}{\bibfnamefont{H.J.}~\bibnamefont{Graffe}} \bibnamefont{et~al.},
  \bibinfo{journal}{Phys.\ Rev.\ Lett.} \textbf{\bibinfo{volume}{96}},
  \bibinfo{pages}{017002} (\bibinfo{year}{2006}).

\bibitem[{\citenamefont{Ofer et~al.}(2006)}]{ofer06}
\bibinfo{author}{\bibfnamefont{R.}~\bibnamefont{Ofer}} \bibnamefont{et~al.},
  \bibinfo{journal}{Phys.\ Rev.\ B} \textbf{\bibinfo{volume}{73}},
  \bibinfo{pages}{012503} (\bibinfo{year}{2006}).

\bibitem[{\citenamefont{Birgeneau et~al.}(2006)}]{birgeneau06}
\bibinfo{author}{\bibfnamefont{R.~J.} \bibnamefont{Birgeneau}}
  \bibnamefont{et~al.} (\bibinfo{year}{2006}).

\bibitem[{\citenamefont{Savici et~al.}(2002)}]{savici02}
\bibinfo{author}{\bibfnamefont{A.~T.} \bibnamefont{Savici}}
  \bibnamefont{et~al.}, \bibinfo{journal}{Phys.\ Rev.\ B}
  \textbf{\bibinfo{volume}{66}}, \bibinfo{pages}{014524}
  (\bibinfo{year}{2002}).

\bibitem[{\citenamefont{Kojima et~al.}(2003)}]{kojima03}
\bibinfo{author}{\bibfnamefont{K.~M.} \bibnamefont{Kojima}}
  \bibnamefont{et~al.}, \bibinfo{journal}{Physica\ B}
  \textbf{\bibinfo{volume}{326}}, \bibinfo{pages}{316} (\bibinfo{year}{2003}).

\bibitem[{\citenamefont{Mohottala et~al.}(2006)}]{mohottala06}
\bibinfo{author}{\bibfnamefont{H.~E.} \bibnamefont{Mohottala}}
  \bibnamefont{et~al.}, \bibinfo{journal}{Nature Physics}
  \textbf{\bibinfo{volume}{5}}, \bibinfo{pages}{377} (\bibinfo{year}{2006}).

\bibitem[{\citenamefont{Alloul et~al.}(1991)}]{alloul91}
\bibinfo{author}{\bibfnamefont{H.}~\bibnamefont{Alloul}} \bibnamefont{et~al.},
  \bibinfo{journal}{Phys.\ Rev.\ Lett.} \textbf{\bibinfo{volume}{67}},
  \bibinfo{pages}{3140} (\bibinfo{year}{1991}).

\bibitem[{\citenamefont{Walstedt}(1993)}]{walstedt93}
\bibinfo{author}{\bibfnamefont{R.~E.} \bibnamefont{Walstedt}} \bibnamefont{et~al.},
  \bibinfo{journal}{Phys.\ Rev.\ B} \textbf{\bibinfo{volume}{48}},
  \bibinfo{pages}{10646} (\bibinfo{year}{1993}).

\bibitem[{\citenamefont{Mahajan et~al.}(1994)}]{mahajan94}
\bibinfo{author}{\bibfnamefont{A.~V.} \bibnamefont{Mahajan}}
  \bibnamefont{et~al.}, \bibinfo{journal}{Phys.\ Rev.\ Lett.}
  \textbf{\bibinfo{volume}{72}}, \bibinfo{pages}{3100} (\bibinfo{year}{1994}).

\bibitem[{\citenamefont{Williams et~al.}(1995)}]{williams95}
\bibinfo{author}{\bibfnamefont{G.~V.~M.} \bibnamefont{Williams}}
  \bibnamefont{et~al.}, \bibinfo{journal}{Phys.\ Rev.\ B}
  \textbf{\bibinfo{volume}{52}}, \bibinfo{pages}{R7034} (\bibinfo{year}{1995}).

\bibitem[{\citenamefont{Williams and Tallon}(1998)}]{williams98}
\bibinfo{author}{\bibfnamefont{G.~V.~M.} \bibnamefont{Williams}}
  \bibnamefont{and} \bibinfo{author}{\bibfnamefont{J.~L.}
  \bibnamefont{Tallon}}, \bibinfo{journal}{Phys.\ Rev.\ B}
  \textbf{\bibinfo{volume}{57}}, \bibinfo{pages}{10984} (\bibinfo{year}{1998}).

\bibitem[{\citenamefont{Bobroff et~al.}(1999)}]{bobroff99}
\bibinfo{author}{\bibfnamefont{J.}~\bibnamefont{Bobroff}} \bibnamefont{et~al.},
  \bibinfo{journal}{Phys.\ Rev.\ Lett.} \textbf{\bibinfo{volume}{83}},
  \bibinfo{pages}{4381} (\bibinfo{year}{1999}).

\bibitem[{\citenamefont{Julien et~al.}(2000)}]{julien00}
\bibinfo{author}{\bibfnamefont{M.-H.} \bibnamefont{Julien}}
  \bibnamefont{et~al.}, \bibinfo{journal}{Phys.\ Rev.\ Lett.}
  \textbf{\bibinfo{volume}{84}}, \bibinfo{pages}{3422} (\bibinfo{year}{2000}).

\bibitem[{\citenamefont{Williams et~al.}(2001)}]{williams01}
\bibinfo{author}{\bibfnamefont{G.}~\bibnamefont{Williams}}
  \bibnamefont{et~al.}, \bibinfo{journal}{Phys.\ Rev.\ B}
  \textbf{\bibinfo{volume}{64}}, \bibinfo{pages}{104506}
  (\bibinfo{year}{2001}).

\bibitem[{\citenamefont{Itoh et~al.}(2003)}]{itoh03}
\bibinfo{author}{\bibfnamefont{Y.}~\bibnamefont{Itoh}} \bibnamefont{et~al.},
  \bibinfo{journal}{Phys.\ Rev.\ B} \textbf{\bibinfo{volume}{67}},
  \bibinfo{pages}{064516} (\bibinfo{year}{2003}).

\bibitem[{\citenamefont{Ouazi et~al.}(2004)}]{ouazi04}
\bibinfo{author}{\bibfnamefont{S.}~\bibnamefont{Ouazi}} \bibnamefont{et~al.},
  \bibinfo{journal}{Phys.\ Rev.\ B} \textbf{\bibinfo{volume}{70}},
  \bibinfo{pages}{104515} (\bibinfo{year}{2004}).

\bibitem[{\citenamefont{Ouazi et~al.}(2006)}]{ouazi06}
\bibinfo{author}{\bibfnamefont{S.}~\bibnamefont{Ouazi}} \bibnamefont{et~al.},
  \bibinfo{journal}{Phys.\ Rev.\ Lett.} \textbf{\bibinfo{volume}{96}},
  \bibinfo{pages}{127005} (\bibinfo{year}{2006}).

\bibitem[{\citenamefont{Nachumi et~al.}(1996)}]{nachumi96}
\bibinfo{author}{\bibfnamefont{B.}~\bibnamefont{Nachumi}} \bibnamefont{et~al.},
  \bibinfo{journal}{Phys.\ Rev.\ Lett.} \textbf{\bibinfo{volume}{77}},
  \bibinfo{pages}{5421} (\bibinfo{year}{1996}).

\bibitem[{\citenamefont{Pan et~al.}(2000)}]{pan00}
\bibinfo{author}{\bibfnamefont{S.~H.} \bibnamefont{Pan}} \bibnamefont{et~al.},
  \bibinfo{journal}{Nature} \textbf{\bibinfo{volume}{403}},
  \bibinfo{pages}{746} (\bibinfo{year}{2000}).

\bibitem[{\citenamefont{Williams
  et~al.}(2005{\natexlab{b}})}]{williams_comment}
\bibinfo{author}{\bibfnamefont{G.}~\bibnamefont{Williams}}
  \bibnamefont{et~al.}, \bibinfo{journal}{Phys.\ Rev.\ B}
  \textbf{\bibinfo{volume}{71}}, \bibinfo{pages}{176502}
  (\bibinfo{year}{2005}{\natexlab{b}}).

\bibitem[{\citenamefont{Bernhard et~al.}(1998{\natexlab{a}})}]{bernhard98_2}
\bibinfo{author}{\bibfnamefont{C.}~\bibnamefont{Bernhard}}
  \bibnamefont{et~al.}, \bibinfo{journal}{Phys.\ Rev.\ Lett.}
  \textbf{\bibinfo{volume}{80}}, \bibinfo{pages}{205}
  (\bibinfo{year}{1998}{\natexlab{a}}).

\bibitem[{\citenamefont{Brian M.~Andersen and Schmid}(2007)}]{andersen07}
\bibinfo{author}{\bibfnamefont{B.M.} \bibnamefont{Andersen},
  \bibfnamefont{P.J.~Hirschfeld}} \bibnamefont{and}
  \bibinfo{author}{\bibfnamefont{M.}~\bibnamefont{Schmid}},
  \bibinfo{journal}{Phys.\ Rev.\ Lett.} \textbf{\bibinfo{volume}{99}},
  \bibinfo{pages}{147002} (\bibinfo{year}{2007}).

\bibitem[{\citenamefont{Proust et~al.}(2002)}]{proust02}
\bibinfo{author}{\bibfnamefont{C.}~\bibnamefont{Proust}} \bibnamefont{et~al.},
  \bibinfo{journal}{Phys.\ Rev.\ Lett.} \textbf{\bibinfo{volume}{89}},
  \bibinfo{pages}{147003} (\bibinfo{year}{2002}).

\bibitem[{\citenamefont{Nakamae et~al.}(2003)}]{nakamae03}
\bibinfo{author}{\bibfnamefont{S.}~\bibnamefont{Nakamae}} \bibnamefont{et~al.},
  \bibinfo{journal}{Phys.\ Rev.\ B} \textbf{\bibinfo{volume}{68}},
  \bibinfo{pages}{100502(R)} (\bibinfo{year}{2003}).

\bibitem[{\citenamefont{Kaminski et~al.}(2003)}]{kaminksi03}
\bibinfo{author}{\bibfnamefont{A.}~\bibnamefont{Kaminski}}
  \bibnamefont{et~al.}, \bibinfo{journal}{Phys.\ Rev.\ Lett.}
  \textbf{\bibinfo{volume}{90}}, \bibinfo{pages}{207003}
  (\bibinfo{year}{2003}).

\bibitem[{\citenamefont{Plate et~al.}(2005)}]{plate05}
\bibinfo{author}{\bibfnamefont{M.}~\bibnamefont{Plate}} \bibnamefont{et~al.},
  \bibinfo{journal}{Phys.\ Rev.\ Lett.} \textbf{\bibinfo{volume}{95}},
  \bibinfo{pages}{077001} (\bibinfo{year}{2005}).

\bibitem[{\citenamefont{Hussey et~al.}(2003)}]{hussey03}
\bibinfo{author}{\bibfnamefont{N.}~\bibnamefont{Hussey}} \bibnamefont{et~al.},
  \bibinfo{journal}{Nature} \textbf{\bibinfo{volume}{425}},
  \bibinfo{pages}{814} (\bibinfo{year}{2003}).

\bibitem[{\citenamefont{Panagopoulos et~al.}(2002)}]{panagopolous02}
\bibinfo{author}{\bibfnamefont{C.}~\bibnamefont{Panagopoulos}}
  \bibnamefont{et~al.}, \bibinfo{journal}{Phys.\ Rev.\ B}
  \textbf{\bibinfo{volume}{66}}, \bibinfo{pages}{064501}
  (\bibinfo{year}{2002}).

\bibitem[{\citenamefont{Kubo et~al.}(1991)}]{kubo91}
\bibinfo{author}{\bibfnamefont{Y.}~\bibnamefont{Kubo}} \bibnamefont{et~al.},
  \bibinfo{journal}{Phys.\ Rev.\ B} \textbf{\bibinfo{volume}{43}},
  \bibinfo{pages}{7875} (\bibinfo{year}{1991}).

\bibitem[{\citenamefont{Oda et~al.}(1991)}]{oda91}
\bibinfo{author}{\bibfnamefont{M.}~\bibnamefont{Oda}} \bibnamefont{et~al.},
  \bibinfo{journal}{Physica\ C} \textbf{\bibinfo{volume}{183}},
  \bibinfo{pages}{234} (\bibinfo{year}{1991}).

\bibitem[{\citenamefont{Nakano et~al.}(1994)}]{nakano94}
\bibinfo{author}{\bibfnamefont{T.}~\bibnamefont{Nakano}} \bibnamefont{et~al.},
  \bibinfo{journal}{Phys.\ Rev.\ B} \textbf{\bibinfo{volume}{49}},
  \bibinfo{pages}{16000} (\bibinfo{year}{1994}).

\bibitem[{\citenamefont{Wakimoto et~al.}(2005)}]{wakimoto05}
\bibinfo{author}{\bibfnamefont{S.}~\bibnamefont{Wakimoto}}
  \bibnamefont{et~al.}, \bibinfo{journal}{Phys.\ Rev.\ B}
  \textbf{\bibinfo{volume}{72}}, \bibinfo{pages}{064521}
  (\bibinfo{year}{2005}).

\bibitem[{\citenamefont{Mashima et~al.}(2006)}]{mashima06}
\bibinfo{author}{\bibfnamefont{H.}~\bibnamefont{Mashima}} \bibnamefont{et~al.},
  \bibinfo{journal}{Phys.\ Rev.\ B} \textbf{\bibinfo{volume}{73}},
  \bibinfo{pages}{060502(R)} (\bibinfo{year}{2006}).

\bibitem[{\citenamefont{Wang et~al.}(2007)}]{wang07}
\bibinfo{author}{\bibfnamefont{Y.}~\bibnamefont{Wang}}
  \bibnamefont{et~al.}, \bibinfo{journal}{Phys.\ Rev.\ B}
  \textbf{\bibinfo{volume}{76}}, \bibinfo{pages}{064512}
  (\bibinfo{year}{2007}).

\bibitem[{\citenamefont{Uemura et~al.}(1993)}]{uemura93}
\bibinfo{author}{\bibfnamefont{Y.}~\bibnamefont{Uemura}} \bibnamefont{et~al.},
  \bibinfo{journal}{Nature} \textbf{\bibinfo{volume}{364}},
  \bibinfo{pages}{605} (\bibinfo{year}{1993}).

\bibitem[{\citenamefont{Schutzmann et~al.}(1994)\citenamefont{Schutzmann,
  Tajima, Miyamoto, and Tanaka}}]{schutzmann94}
\bibinfo{author}{\bibfnamefont{J.}~\bibnamefont{Schutzmann}},
  \bibinfo{author}{\bibfnamefont{S.}~\bibnamefont{Tajima}},
  \bibinfo{author}{\bibfnamefont{S.}~\bibnamefont{Miyamoto}}, \bibnamefont{and}
  \bibinfo{author}{\bibfnamefont{S.}~\bibnamefont{Tanaka}},
  \bibinfo{journal}{Phys.\ Rev.\ Lett.} \textbf{\bibinfo{volume}{73}},
  \bibinfo{pages}{174} (\bibinfo{year}{1994}).

\bibitem[{\citenamefont{Uchida et~al.}(1996)\citenamefont{Uchida, Tamasaki, and
  Tajima}}]{uchida96}
\bibinfo{author}{\bibfnamefont{S.}~\bibnamefont{Uchida}},
  \bibinfo{author}{\bibfnamefont{K.}~\bibnamefont{Tamasaku}}, \bibnamefont{and}
  \bibinfo{author}{\bibfnamefont{S.}~\bibnamefont{Tajima}},
  \bibinfo{journal}{Phys.\ Rev.\ B} \textbf{\bibinfo{volume}{53}},
  \bibinfo{pages}{14558} (\bibinfo{year}{1996}).

\bibitem[{\citenamefont{Wen et~al.}(2000)\citenamefont{Wen, Chen, Yang, and
  Zhao}}]{wen00}
\bibinfo{author}{\bibfnamefont{H.H.}~\bibnamefont{Wen}},
  \bibinfo{author}{\bibfnamefont{X.}~\bibnamefont{Chen}},
  \bibinfo{author}{\bibfnamefont{W.}~\bibnamefont{Yang}}, \bibnamefont{and}
  \bibinfo{author}{\bibfnamefont{Z.}~\bibnamefont{Zhao}},
  \bibinfo{journal}{Phys.\ Rev.\ Lett.} \textbf{\bibinfo{volume}{85}},
  \bibinfo{pages}{2805} (\bibinfo{year}{2000}).

\bibitem[{\citenamefont{Wakimoto et~al.}(2004)}]{wakimoto04}
\bibinfo{author}{\bibfnamefont{S.}~\bibnamefont{Wakimoto}}
  \bibnamefont{et~al.}, \bibinfo{journal}{Phys.\ Rev.\ Lett.}
  \textbf{\bibinfo{volume}{92}}, \bibinfo{pages}{217004}
  (\bibinfo{year}{2004}).

\bibitem[{\citenamefont{Savici et~al.}(2005)}]{savici05}
\bibinfo{author}{\bibfnamefont{A.~T.} \bibnamefont{Savici}}
  \bibnamefont{et~al.}, \bibinfo{journal}{Phys.\ Rev.\ Lett.}
  \textbf{\bibinfo{volume}{95}}, \bibinfo{pages}{157001}
  (\bibinfo{year}{2005}).

\bibitem[{\citenamefont{Ishida et~al.}(2007)}]{ishida07}
\bibinfo{author}{\bibfnamefont{K.}~\bibnamefont{Ishida}} \bibnamefont{et~al.},
  \bibinfo{journal}{J.\ Magn.\ Magn.\ Mater.} \textbf{\bibinfo{volume}{310}},
  \bibinfo{pages}{526} (\bibinfo{year}{2007}).

\bibitem[{\citenamefont{Sonier et~al.}(2008)}]{sonier08}
\bibinfo{author}{\bibfnamefont{J.}~\bibnamefont{Sonier}} \bibnamefont{et~al.},
  \bibinfo{journal}{Phys.\ Rev.\ Lett.} \textbf{\bibinfo{volume}{101}},
  \bibinfo{pages}{117001} (\bibinfo{year}{2008}).

\bibitem[{\citenamefont{MacDougall et~al.}(2006)}]{macdougall06}
\bibinfo{author}{\bibfnamefont{G.~J.} \bibnamefont{MacDougall}}
  \bibnamefont{et~al.}, \bibinfo{journal}{Physica B}
  \textbf{\bibinfo{volume}{374-375}}, \bibinfo{pages}{211}
  (\bibinfo{year}{2006}).

\bibitem[{\citenamefont{Lee et~al.}(1998)\citenamefont{Lee, Kilcoyne, and
  Cywinksi}}]{summerschool}
\bibinfo{editor}{\bibfnamefont{S.}~\bibnamefont{Lee}},
  \bibinfo{editor}{\bibfnamefont{S.}~\bibnamefont{Kilcoyne}}, \bibnamefont{and}
  \bibinfo{editor}{\bibfnamefont{R.}~\bibnamefont{Cywinksi}}, eds.
  (\bibinfo{publisher}{Scottish Universities Summer School in Physics and
  Institute of Physics Publishing, Bristol and Philadelphia},
  \bibinfo{year}{1998}).

\bibitem[{\citenamefont{MacDougall et~al.}(2008)}]{macdougall08}
\bibinfo{author}{\bibfnamefont{G.J.}~\bibnamefont{MacDougall}}
  \bibnamefont{et~al.}, \bibinfo{journal}{Phys.\ Rev.\ Lett.}
  \textbf{\bibinfo{volume}{101}}, \bibinfo{pages}{017001} (\bibinfo{year}{2008}).

\bibitem[{\citenamefont{Zhang and Rice}(1988)}]{zhang88}
\bibinfo{author}{\bibfnamefont{F.C.}~\bibnamefont{Zhang}} \bibnamefont{and}
  \bibinfo{author}{\bibfnamefont{T.}~\bibnamefont{Rice}},
  \bibinfo{journal}{Phys.\ Rev.\ B} \textbf{\bibinfo{volume}{37}},
  \bibinfo{pages}{3759} (\bibinfo{year}{1988}).

\bibitem[{\citenamefont{Eskes and Sawatzky}(1988)}]{eskes88}
\bibinfo{author}{\bibfnamefont{H.}~\bibnamefont{Eskes}} \bibnamefont{and}
  \bibinfo{author}{\bibfnamefont{G.A.}~\bibnamefont{Sawatzky}},
  \bibinfo{journal}{Phys.\ Rev.\ Lett.} \textbf{\bibinfo{volume}{61}},
  \bibinfo{pages}{1415} (\bibinfo{year}{1988}).

\bibitem[{\citenamefont{Eskes and Sawatzky}(1991)}]{eskes91}
\bibinfo{author}{\bibfnamefont{H.}~\bibnamefont{Eskes}} \bibnamefont{and}
  \bibinfo{author}{\bibfnamefont{G.A.}~\bibnamefont{Sawatzky}},
  \bibinfo{journal}{Phys.\ Rev.\ B} \textbf{\bibinfo{volume}{44}},
  \bibinfo{pages}{9656} (\bibinfo{year}{1991}).

\bibitem[{\citenamefont{Maignan et~al.}(1990)}]{maignan90}
\bibinfo{author}{\bibfnamefont{A.}~\bibnamefont{Maignan}} \bibnamefont{et~al.},
  \bibinfo{journal}{Physica C} \textbf{\bibinfo{volume}{170}},
  \bibinfo{pages}{350} (\bibinfo{year}{1990}).

\bibitem[{\citenamefont{Martin et~al.}(1995)}]{martin95}
\bibinfo{author}{\bibfnamefont{C.}~\bibnamefont{Martin}} \bibnamefont{et~al.},
  \bibinfo{journal}{Chem. Mater.} \textbf{\bibinfo{volume}{7}},
  \bibinfo{pages}{1414} (\bibinfo{year}{1995}).

\bibitem[{\citenamefont{Haskel et~al.}(1997)}]{haskel97}
\bibinfo{author}{\bibfnamefont{D.}~\bibnamefont{Haskel}} \bibnamefont{et~al.},
  \bibinfo{journal}{Phys.\ Rev.\ B} \textbf{\bibinfo{volume}{56}},
  \bibinfo{pages}{R521} (\bibinfo{year}{1997}).

\bibitem[{\citenamefont{Chen et~al.}(1992)}]{chen92}
\bibinfo{author}{\bibfnamefont{C.}~\bibnamefont{Chen}} \bibnamefont{et~al.},
  \bibinfo{journal}{Phys.\ Rev.\ Lett.} \textbf{\bibinfo{volume}{68}},
  \bibinfo{pages}{2543} (\bibinfo{year}{1992}).

\bibitem[{\citenamefont{Pellegrin et~al.}(1993)}]{pellegrin93}
\bibinfo{author}{\bibfnamefont{E.}~\bibnamefont{Pellegrin}}
  \bibnamefont{et~al.}, \bibinfo{journal}{Phys.\ Rev.\ B}
  \textbf{\bibinfo{volume}{47}}, \bibinfo{pages}{3354} (\bibinfo{year}{1993}).

\bibitem[{\citenamefont{Srivastava et~al.}(1996)}]{srivastava96}
\bibinfo{author}{\bibfnamefont{P.}~\bibnamefont{Srivastava}}
  \bibnamefont{et~al.}, \bibinfo{journal}{Phys.\ Rev.\ B}
  \textbf{\bibinfo{volume}{54}}, \bibinfo{pages}{693} (\bibinfo{year}{1996}).

\bibitem[{\citenamefont{Kopp et~al.}(2007)\citenamefont{Kopp, Ghosal, and
  Chakravarty}}]{kopp07}
\bibinfo{author}{\bibfnamefont{A.}~\bibnamefont{Kopp}},
  \bibinfo{author}{\bibfnamefont{A.}~\bibnamefont{Ghosal}}, \bibnamefont{and}
  \bibinfo{author}{\bibfnamefont{S.}~\bibnamefont{Chakravarty}},
  \bibinfo{journal}{Proc.\ Natl.\ Acad.\ Sci.\ USA}
  \textbf{\bibinfo{volume}{104}}, \bibinfo{pages}{6123} (\bibinfo{year}{2007}).

\bibitem[{\citenamefont{Barbiellini and Jarlborg}(2008)}]{barbiellini08}
\bibinfo{author}{\bibfnamefont{B.}~\bibnamefont{Barbiellini}} \bibnamefont{and}
  \bibinfo{author}{\bibfnamefont{T.}~\bibnamefont{Jarlborg}},
  \bibinfo{journal}{Phys.\ Rev.\ Lett.} \textbf{\bibinfo{volume}{101}},
  \bibinfo{pages}{157002} (\bibinfo{year}{2008}).

\bibitem[{\citenamefont{Mendels et~al.}(1994)}]{mendels94}
\bibinfo{author}{\bibfnamefont{P.}~\bibnamefont{Mendels}} \bibnamefont{et~al.},
  \bibinfo{journal}{Phys.\ Rev.\ B} \textbf{\bibinfo{volume}{49}},
  \bibinfo{pages}{10035} (\bibinfo{year}{1994}).

\bibitem[{\citenamefont{Bernhard et~al.}(1998{\natexlab{b}})}]{bernhard98}
\bibinfo{author}{\bibfnamefont{C.}~\bibnamefont{Bernhard}}
  \bibnamefont{et~al.}, \bibinfo{journal}{Phys.\ Rev.\ B}
  \textbf{\bibinfo{volume}{58}}, \bibinfo{pages}{R8937}
  (\bibinfo{year}{1998}{\natexlab{b}}).

\bibitem[{\citenamefont{Mendels et~al.}(1999)}]{mendels99}
\bibinfo{author}{\bibfnamefont{P.}~\bibnamefont{Mendels}} \bibnamefont{et~al.},
  \bibinfo{journal}{Europhys.\ Lett.} \textbf{\bibinfo{volume}{46}},
  \bibinfo{pages}{678} (\bibinfo{year}{1999}).

\bibitem[{\citenamefont{Gabay et~al.}(2007)}]{gabay07}
\bibinfo{author}{\bibfnamefont{M.}~\bibnamefont{Gabay}} \bibnamefont{et~al.},
  \bibinfo{journal}{Phys.\ Rev.\ B} \textbf{\bibinfo{volume}{77}}, \bibinfo{pages}{165110}
  (\bibinfo{year}{2008}).

\bibitem[{\citenamefont{Risdiana et~al.}(2008)}]{risdiana08}
\bibinfo{author}{\bibnamefont{Risdiana}} \bibnamefont{et~al.},
  \bibinfo{journal}{Phys.\ Rev.\ B} \textbf{\bibinfo{volume}{77}},
  \bibinfo{pages}{054516} (\bibinfo{year}{2008}).

\bibitem[{\citenamefont{Lee et~al.}(2004)}]{lee04}
\bibinfo{author}{\bibfnamefont{Y.~S.} \bibnamefont{Lee}} \bibnamefont{et~al.},
  \bibinfo{journal}{Phys.\ Rev.\ B} \textbf{\bibinfo{volume}{69}},
  \bibinfo{pages}{020502(R)} (\bibinfo{year}{2004}).

\end{thebibliography}
\end{document}